\documentclass[10pt]{article}
\usepackage{epsfig}
\usepackage{amsfonts}
\usepackage{graphicx}
\usepackage{epsfig}

\usepackage{amsmath}

\date{\today}

\newcommand{\la}{\lambda}
\newcommand{\f}{\phi}
\newcommand{\ka}{\kappa}

\newcommand{\ee}{\end{equation}}
\newcommand{\eea}{\end{eqnarray}}
\newcommand{\be}{\begin{equation}}
\newcommand{\bea}{\begin{eqnarray}}

\newcommand{\vep}{\varepsilon}

\newcommand{\re}[1]{(\ref{#1})}

\begin{document}

\title{ Asymptotic analysis of the Skyrmed monopole}
\author{{\large Yves Brihaye,}$^{\ddagger}$ {\large J{\" u}rgen Burzlaff
}$^{\diamond \star}$
and {\large D. H. Tchrakian}$^{\dagger \star}$
\\
\\
$^{\ddagger}${\small Physique-Math\'ematique, Universite de
Mons-Hainaut, Mons, Belgium}
\\
$^{\diamond}${\small School of Mathematical Sciences, Dublin City
University, Dublin 9, Ireland}
\\
$^{\dagger}${\small Department of
Mathematical Physics, National University of Ireland Maynooth,}
{\small Maynooth, Ireland}
\\
$^{\star}${\small School of Theoretical Physics -- DIAS, 10
Burlington
Road, Dublin 4, Ireland }}

\maketitle

\begin{abstract}
We consider a variant of the Georgi Glashow model in the BPS limit,
augmented by a higher derivative Skyrme-like term, which is the simplest
YMH model that can support monopole bound states. The spherically symetric
solutions are studied with a combination of analytic and numerical
techniques, which strongly suggest that the solutions converge to a
finite energy configuration in the limit of infinite coupling of the
Skyrme-like term.

\end{abstract}

\section{Introduction}

The asymptotic analysis for the {\it unit} charge 't~Hooft-Polyakov
monopole~\cite{'t Hooft:1974qc,Polyakov:1974ek} was carried out long ago by Kirkman and
Zachos~\cite{Kirkman:1981ck} and by Gardner~\cite{Gardner:1982fk}, and, has recently been elaborated
by Forg\'acs {\it et. al.}~\cite{Forgacs:2005vx}
by providing a high precision numerical analysis of the problem. In contrast to the Prasad-Sommerfield
monopole~\cite{Prasad:1975kr} which
is evaluated in closed form, the 't~Hooft-Polyakov monopole can be evaluated only numerically
since it is a solution to the Georgi-Glashow model that
exhibits a symmetry breaking Higgs self interaction potential. The existence of this numerically
evaluated solution is underpinned by the purely analytic proof of existence given by Tyupkin,
Fate'ev and Schwarz~\cite{Tyupkin:1975pk}, but this does not shed any light on the behaviour of the
solution as a function of the strength $\la$ of the Higgs potential term. Numerical studies of this $\la$ dependence,
in \cite{BM} for the spherically symmetric case and in \cite{Kleihaus:1998wt} for the axially symmetric case,
reveal in particular that as $\la\to\infty$ the energy of the monopole asymptotes to a finite value. It is
this behaviour on $\la$ found numerically that is underpinned by the analytic analysis of
\cite{Kirkman:1981ck}.

It is our intention in this short note to supply an asymptotic
analysis similar to that of \cite{Kirkman:1981ck}, for the {\it
unit} charge monopole of a variant of the Georgi-Glashow model
characterised by the addition of a Skyrme like term in terms of
the covariant detivatives of the Higgs field. The model is the
$SO(3)$ Higgs model \be \label{H} {\cal
H}=\frac18\left(\frac14|F_{ij}^{ab}|^2+|D_i\phi^a|^2+
\frac14\ka|D_{[i}\phi^aD_{j]}\phi^b|^2\right)\ge
4\pi\varrho_1\,, \ee $\ka$ giving the strength of the coupling of
the Skyrme like term, and $\ka$ having the dimension of $L^4$. The
lower bound on the right hand side is the usual monopole charge
density \be \label{top}
\varrho_1=\frac{1}{16}\vep_{aa'b}\vep^{ijk}F_{ij}^{aa'}D_k\f^b \ee
featuring a Skyrme like term {\it in lieu} of the Higgs symmetry
breaking potential in the Georgi-Glashow model. This monopole,
which was described in \cite{Grigoriev:2002qc} as a Skyrmed
monopole, is a solution to a model that was distilled from a
rather more involved set of models
\cite{Kleihaus:1998kd,Kleihaus:1998gy}, the latter being designed
to support monolpoles with both mutually attracting and repelling
phases. (The Skyrmed monopole of \cite{Grigoriev:2002qc} supports
bound states which are axially symmetric for charges higher than
$2$, in contrast to Skyrmions of charges $3$ and higher which
instead exhibit Platonic symmetries~\cite{Battye:2000se}.)

In contrast to the 't~Hooft-Polyakov monopole for which there is
an analytic proof of existence~\cite{Tyupkin:1975pk}, there are no
such proofs for the monopoles of the various generalised
models~\cite{Kleihaus:1998kd,Kleihaus:1998gy,Grigoriev:2002qc}.
This is because of the presence of higher order Skyrme like terms,
and the only known solutions are those constructed numerically.
Here we will supply an asymptotic analysis analogous to that of
\cite{Kirkman:1981ck}, for the {\it unit} charged Skyrmed
monopole~\cite{Grigoriev:2002qc} which is the simplest such
example available. Due to the considerably more complex structure
of the equations here, it is very difficult to perform the
promised asymptotic analysis using purely analytic method, and
intead we present a combination of both analytic and numerical
analysis. (Similar techniques were used in \cite{Brihaye:2004pz},
in the context of the $SO(3)$ gauged
Skyrmion~\cite{Arthur:1996wy,BT}.)

In the next two sections we present the asymptotic and the numerical analyses, respectively, followed by
a brief summary of our result.

\section{Asymptotic analysis}

Since we are restricting to the charge-$1$ monopole of \re{H}, we
impose the usual spherically symmetric Ansatz \be \label{sph}
A_i^{[aa']}=\frac{1-w(r)}{r}\,\hat x^{[a}\,\delta^{a']}\, ,\quad
\f^a=\eta\,h(r)\,\hat x^{a} \ee where the brackets $[ab]$ imply
antisymmetrisation, $\hat x^{a}$ is the unit position vector and
$\eta$ is the dimensionful VEV of the Higgs field. Imposing
\re{sph}, the residual one dimensional static Hamiltonian is \be
\label{redH} H = w'^2 + \frac{(w^2-1)^2}{2 r^2} +
\frac12\,r^2\,h'^2 + w^2\,h^2 +
\kappa\,w^2\,h^2\left(2h'^2+\frac{w^2h^2}{r^2}\right)\,, \ee
having rescaled $r\to\eta r$ and $\kappa\to\eta\ka$ so that both
the rescaled radial variable $r$ and the rescaled Skyrme coupling
$\ka$ in \re{redH} are dimensionless. The corresponding equations
for $w$ and $h$ are
\[ w'' + \frac{w(1-w^2 )}{r^2} - wh^2
= 2 \kappa wh^2 \left( (h')^2 + \frac{w^2 h^2}{r^2} \right)\, ,
\]
\[ r^2 h'' + 2rh' - 2w^2 h \]
\be\label{eqs} = 4\kappa w^2 h \left( (h')^2 + \frac{w^2 h^2}{r^2} \right)
-4\kappa \left( w^2 h^2 h'' + 2w^2 h (h')^2 + 2 ww' h^2 h'
\right) \ee

Due to their complex structure we are only able to extract from
the equations some asymptotic information near the origin, at
infinity, and for $\kappa\rightarrow\infty$. The asymptotic
analysis will then be complemented by the numerical results. For
$r\rightarrow 0$, we have $w\rightarrow 1$ and $h\rightarrow 0$. A
dominant balance analysis gives
\[ w = 1 + w_2 r^2 + w_4 r^4 + O(r^6) \, ,
\quad h = h_1 r + h_3 r^3 + O(r^5)\] By induction we see that the
asymptotic expansions contain only even or odd powers of $r$ for
$w$ and $h$, respectively. For $\kappa =0$, the Prasad-Sommerfield
solution yields $w_2 =-\frac{1}{6}$ and $h_1 = \frac{1}{3}$. For
nonzero $\kappa$, $w_2$ and $h_1$ have to be determined
numerically. The
coefficients for the next highest order are
\[ w_4 = \frac{3}{10} w_2^2 + \frac{1}{10} (1 + 4 \kappa h_1^2
) h_1^4 \, , \quad h_3 = \frac{2-4\kappa h_1^2}{5+20 \kappa
h_1^2}\; w_2h_1 \, , \] and all other coefficients can be
calculated recursively.

For $r\rightarrow\infty$, we have $w\rightarrow 0$ and
$h\rightarrow 1$. Here the dominant balance analysis leads to an
exponential fall-off for $w$ and to
\[ r^2 h'' + 2rh' =0 \]
for the leading term in $h$. We therefore have
\be
\label{asymptotic}
 h=1-\frac{q}{r} + O\left( \alpha (r) e^{-2r} \right)\, , \quad
w=\beta (r)e^{-r} + O\left( \gamma (r) e^{-2r} \right) 
\ee
 Using
again induction, we see that the asymptotic expansion at infinity
for $h$ contains only even powers of $e^{-r}$, whereas the
asymptotic expansion for $w$ contains the odd powers of $e^{-r}$.
The coefficient functions in front of the exponential functions,
starting with $\alpha (r)$ and $\beta (r)$, are polynomially
bounded. For $\kappa =0$, we have $q=1$ and $\beta =2r$. For
nonzero $\kappa$, $\beta (r)$ satisfies the equation
\[ \beta ''(r) -2\beta '(r) +\frac{1}{r^2} \beta (r) +
\left(\frac{2q}{r} - \frac{q^2}{r^2}\right) \beta (r) = 2\kappa
\left(\frac{q^2}{r^4} - \frac{2 q^3}{r^5} + \frac{q^4}{r^6}\right)
\beta (r) \] and therefore
\[ \beta (r) = p\; r^q \left( 1 + \frac{q-1}{2r} + O
\left(\frac{1}{r^2}\right) \right) \]
\section{Dependance on $\kappa$}
To study the dependence of the energy
\[ E= \int_0^\infty H \; dr \]
on $\kappa$, we calculate
\[
\frac{dE}{d\kappa} = \int_0^\infty \left( \frac{\partial
H}{\partial\kappa} + \frac{\partial w}{\partial \kappa
^2} \frac{\partial H}{\partial w} + \frac{\partial
w'}{\partial \kappa}\frac{\partial H}{\partial w'} +
\frac{\partial h}{\partial \kappa}\frac{\partial H}{\partial h}
+ \frac{\partial h'}{\partial \kappa
^2}\frac{\partial H}{\partial h'} \right) dr \]
\be
\label{Esk} = \int_0^\infty w^2 h^2 \left( 2(h')^2
+ \frac{w^2 h^2}{r^2} \right) dr \stackrel{\rm def.}= E_{sk} > 0 \ee 
Here we have used
integration by parts, the equations for $w$ and $h$, and the
boundary conditions. We see that the energy increases with
$\kappa$. We also see that, if the energy is bounded as
$\kappa\rightarrow\infty$ (as suggested by the numerical results reported in 
the next section), $E_{sk}$ and therefore the product $wh$
must vanish in this limit. Because of the boundary conditions, $w$ cannot
be zero for small $r$, whereas $h$ cannot vanish for large $r$. This
strongly suggests that $h$ is zero in some interval $(0,r_m)$
and that $w$ is zero in the interval $(r_m ,\infty )$.

In the limit $\kappa\rightarrow\infty$ we therefore expect the following equations to hold,
\begin{equation}\label{hzero} w''(r) +
\frac{w(r)(1-w^2 (r) )}{r^2} = 0 \, ,\quad h(r)=0 \quad (0<r<r_m ) ,
\end{equation}
\begin{equation}\label{wzero} w(r)=0\, , \quad r^2 h''(r) + 2rh'(r) =0 \quad (r_m
<r<\infty ) \end{equation} with boundary conditions
\[ w(0) =1\, , \quad w(r_m )=0\, ,\quad h(r_m)=0\quad {\rm and}\quad h\rightarrow 1
\;\; {\rm as} \;\; r\rightarrow\infty \] The solution of equation \re{wzero} is
\[ h(r) =0\quad (0<r<r_m )\, , \quad h(r)=1-\frac{r_m}{r} \quad (r_m
<r<\infty ) \] The solutions to equation \re{hzero} have been studied a long time ago \cite{Lemke}, but not to the same extent as other special functions. One property we can deduce immediately from equation \re{hzero} is that, because $w''$ is negative for $r<r_m$, $w'(r_m )$ cannot be zero. 

Denoting the solution of Eqs. (\ref{hzero}),(\ref{wzero}) with the
appropriate boundary condition at $r=r_m$ by $w_{\infty},h_{\infty}$,
the corresponding energy can be obtained easily~:
\be
\label{Einfty} 
E_\infty (r_m) = \int_0^{r_m} \left(w_{\infty}'^2 + \frac{(w_{\infty}^2-1)^2}{2
r^2}\right) dr + \int_{r_m}^\infty \left( \frac{1}{2r^2} +
\frac12\,r^2\,h_{\infty}'^2 \right) dr = E_w (r_m ) + \frac{1+r_m^2}{2r_m} \ \ .
\ee
From the second term we see that $E_\infty \rightarrow \infty$ for $r_m \rightarrow 0$ and for $r_m \rightarrow \infty$; so $E_\infty (r_m )$ has a minimum. 
Solving Eq. \re{hzero} numerically for several values of $r_m$ and computing the value 
$E_\infty (r_m)$, we have determined the local minimum of $E_{\infty}$ which we find to occur for $r_{m,c} \approx 2.0623$. 
Furthermore, our result strongly suggests the relation $E_\infty (r_{m,c}) = r_{m,c}$;
we have no analytic proof for this to be an identity but it 
 holds whithin our numerical
accuracy i.e. $10^{-4}$. The configuration minimizing $E_\infty$ has $w_{\infty}''(0) \approx - 0.656$.

The results of the numerical analysis reported in the
next section will strongly confirm that 
\be
     \lim_{\kappa \to \infty} (w(r),h(r)) = (w_{\infty},h_{\infty}) \ \  {\rm with} \ \  r_m = r_{m,c}.
\ee 
for the solution $w,h$  of Eqs. (\ref{eqs}). 
\section{Numerical analysis}

\begin{figure}
\centering
\epsfysize=8cm
\mbox{\epsffile{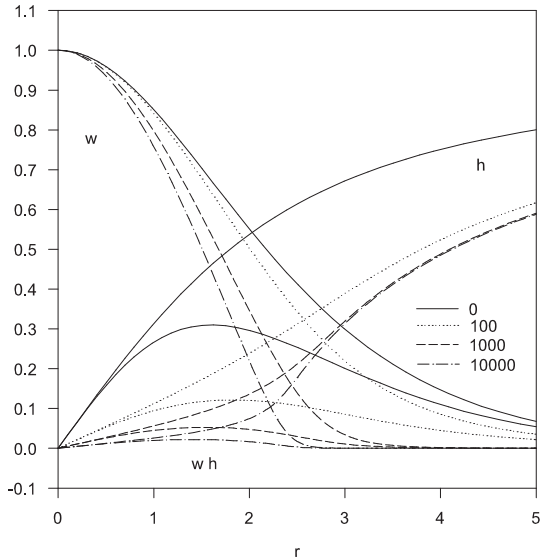}}
\caption{\label{fig1}
The profile for the solution for several values of $\kappa$.}
\end{figure}

We now discuss the numerical solutions of Eq. (\ref{eqs}) for finite $\kappa$.
In the limit $\kappa=0$, the classical equations coincide with the equations
of the  BPS monopole which is known explicitly.
To the best of our knowledge, no explicit solution exist for $\kappa > 0$. We have solved 
equations \re{eqs} completed with the boundary conditions,
\be
     w(0) = 1 \  \ , \ \ h(0) = 0 \ \ , \ \ w(\infty)=0 \ \ , \ \ h(\infty) = 1
\ee  
by using a numerical solver \cite{COLSYS}.
The PBS monopole gets smoothly deformed for $\kappa > 0$. This is illustrated in Fig. \ref{fig1},
where the profiles $w,h$ of the BPS monopole solution are superposed
with solutions corresponding to several positive values of $\kappa$. For reasons explained 
in the previous section, we supplemented
this figure with the profiles of the product $w h$.

\begin{figure}
\centering
\epsfysize=8cm
\mbox{\epsffile{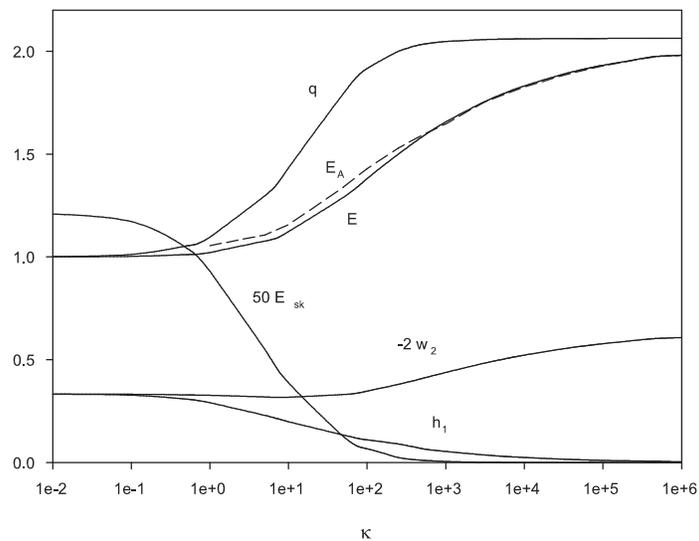}}
\caption{\label{fig2}
The values of $E,E_{sk},q, h_1\equiv h'(0),w_2 \equiv w''(0)/2$ are plotted as functions of $\kappa$ (solid lines)
the values of $E_A$ are represented with the dashed line.}
\end{figure}

Several parameters characterizing the solutions, namely the classical energy $E$, the energy of the Skyrme
term $E_{sk}$ (see \re{Esk}), the value of $q$ (see (\ref{asymptotic})), 
$h_1 \equiv h'(0)$ and $w_2 \equiv w''(0)/2$ are plotted as functions
of $\kappa$ in Fig.
\ref{fig2}.
The characteristics of the PBS monopole are recovered in the limit $\kappa =0$, for instance
$E=1$, $q=1$, $h'(0)=1/3$, $w''(0) = -1/6$.
\begin{figure}
\centering
\epsfysize=8cm
\mbox{\epsffile{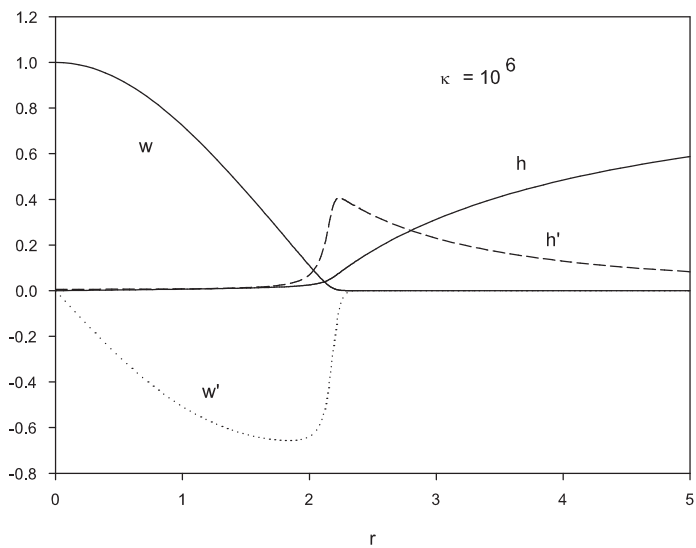}}
\caption{\label{fig3}
The profile for the solution corresponding to $\kappa=10^6$ }
\end{figure}
The natural challenge is  to construct numerically the solutions for large values of $\kappa$
and to confirm that they  evolve according to the pattern discussed in the previous section.
The  quantities $w_2,h_1,q,E$  extracted from our numerical solutions
are reported on Fig. \ref{fig2}. The figure shows that they stay  finite
for $\kappa \gg 1$.
In particular for $\kappa \geq 10^6$ we find $ q \approx 2.0623$, $w''(0) \approx - 0.656$, $h(0) \approx 0$
  suggesting that, in the limit $\kappa \to \infty$,
the solution   approaches the configuration (\ref{hzero}),(\ref{wzero}) which minimizes the energy $E_{\infty}$.
Our numerical results further indicate that  the  
 the product $w h$ tends uniformly to the null function
 in the limit $\kappa \to \infty$.
As anticipated in the previous section, we observe from our numerical solutions
that, in the interior region, i.e. for $r \in [0,r_m]$  we have $h(r) \sim 0$, while in the exterior region we have rather  $w(r) \sim 0$. 
Of course the value of $r_m$ cannot be precisely determined for $\kappa < \infty$ but
the numerical results obtained for large $\kappa$
are quite compatible with $r_m = r_{m,c}  \approx 2.0623$ and with $r_m = q$.
This is illustrated in Fig. \ref{fig3} for $\kappa = 10^6$ where the functions $w,h$ and the derivatives $w',h'$
are plotted as functions of $r$. 
The limit of the solutions of the equations \re{eqs} is therefore not differentiable at $r=r_m$, i.e., the equations \re{eqs} are singularly perturbed about $1/\kappa =0$.
As a consequence of the absence of differentiability of the limiting solution at an intermediate
point of the domain of integration,
the numerical analysis becomes  involved for increasing $\kappa$.
%

Coming finally to the energy of the solution, the numerical evaluation of $E$
is fully compatible with the fact that the energy stays finite and that 
$E_{\kappa \to \infty}=2.0623$. The convergence of $E$ for $\kappa \to \infty$ is, however,
much slower than the convergence of the parameter $q$. 
To argue that this statement is correct, we  further evaluate the quantity
\be
 \label{EA} 
E_A (\kappa) = \int_0^{q} \left(w'^2 + \frac{(w^2-1)^2}{2
r^2}\right) dr     +  \frac{1+q^2}{2q} \ \ . 
\ee
with the numerical profile of $w(r)$.
In fact, $E_A$ is an estimation of the energy obtained by assuming 
$h=1 - q/r$, $w=0$ for $r\in [q,\infty]$ and $h=0$ and the numerical value of $w(r)$   for  $r \in [0,q]$.
One should expect $E_A(\kappa \to \infty) = E_{\infty}(r_{m,c})$ since $q \sim r_{m,c}$ in this limit.

 The value $E_A$ is reported on Fig. \ref{fig2}  (see the dashed line);
confirming our expectation, we find that the difference $E-E_A$ tends quickly to zero (in fact exponentially fast)
for $\kappa \to \infty$.
Interestingly,  $E_A$ provides a reasonably good approximation to the energy $E$ even for
small values of $\kappa$; for instance we find $E_A/E \sim 1.035$ for $\kappa=1$.

\section{Summary}

We have carried out a combination of analytic and numerical analysis for what we refer to as the Skyrmed
monopole, which is a finite energy solution to the equations of the model \re{H}.
The asymptotic analysis involves the study of a one parameter family of monopole solutions, parametrised by
the strength of the coupling of the quartic Skyrme like term. For any value of the coupling constant $\kappa$ we have given asymptotic expansions of the solutions for $r \rightarrow 0$ and $r \rightarrow \infty$. We have concentrated, however, mainly on the behaviour of the solutions for $\kappa\rightarrow\infty$. The numerical analysis shows that the energy is bounded as $\kappa\rightarrow\infty$, a result for which we have no mathematical underpinning. Given that the energy is bounded, we can however deduce analytically some interesting results. We find in particular that the equations \re{eqs} are singularly perturbed about $1/\kappa =0$. This is reminiscent of Burgers' equation with a small coefficient in front of the second order derivative. When this coefficient goes to zero, the solutions of Burgers' equations tend to a weak solution with a shock, i.e., the limiting weak solution is discontinuous. In our case, the derivative of the limiting weak solution is discontinuous. This analytic result, and the results we get for the values of some typical constants are all supported by our numerical analysis.


\bigskip
\noindent
{\bf Acknowledgement}

\medskip
We thank Cosmas Zachos for helpful comments on this work.
This work was carried out in the framework of Science Foundation Ireland
(SFI) Research Frontiers Programme (RFP) project RFP07/FPHY330.
YB is grateful to the Belgian FNRS for financial support.




\end{document}